\documentclass[preprint,12pt]{revtex4}

 \usepackage{xcolor}
\usepackage{graphicx}
\usepackage{amssymb}
\usepackage{cancel}
\usepackage{ulem}
\usepackage{lineno}
\usepackage{amsmath}
\usepackage{appendix}

\newcommand{\orcid}[1]{\href{https://orcid.org/#1}
{\includegraphics[width=7pt]{orcid.eps}}}

\begin{document}

\title{
Superstatistical approach of the anomalous exponent   \\ for scaled Brownian motion 
}

\author{M. A. F. dos Santos$^{1}$, L. Menon Jr.$^{1}$ 
 and D. Cius$^{2}$}
\address{$^{1}$ Department of Physics, PUC-Rio, Rua Marqu\^es de S\~ao Vicente 225, 22451-900, Rio de Janeiro, RJ, Brazil}
\address{$^{2}$ Programa de P\'os-Gradua\c{c}\~{a}o Ci\^{e}ncias/F\'{i}sica,
  Universidade Estadual de Ponta Grossa,
  84030-900 Ponta Grossa, Paran\'a, Brazil}

\begin{abstract}

Anomalous diffusion phenomenon is an intriguing process that tracer diffusion presents in numerous complex systems. Current experimental and theoretical investigations have reported the emergence of random diffusivity scenarios accompanied by the heterogeneity of the $\alpha$-anomalous diffusion exponents.
In this framework, we investigate a heterogeneous ensemble of tracers governed by scaled Brownian motion (sBm). The heterogeneous features are considered on anomalous diffusion exponent and diffusivity. We introduce two superstatistics of anomalous exponent, the truncated-Gaussian and truncated $\chi^2$-Gamma distributions. 
In this way, we show how the statistics of anomalous exponent affect the spreading of a mixture of particles governed by sBm.
  We also analyse the role of different temporal scales of sBm in superstatistics. Furthermore, we investigate the effects of coupling between diffusivity and anomalous exponent on superstatistics of sBm.
 The investigation provides a thorough analysis of simulation and analytical approach.
 The results imply rich classes of anomalous diffusion processes accompanied by non-Gaussian diffusion. \\

\end{abstract}

  

\maketitle

 
\section{Introduction} \label{sec:intro}

The Brownian motion (Bm) was firstly observed in the erratic motion of tiny particles in a simple fluid \cite{brown1828mikroskopische,brown1828}. After that, it was modelled by different ways \cite{einstein1905molekularkinetischen,sutherland1905lxxv,von1906kinetischen,Langevin1908} that were experimentally validated by Perrin  \cite{perrin1908agitation,perrin1909}. Since then, it is known that the tracer diffusion of Brownian particles has two remarkable features: $(i)$ linear mean square displacement at $t$-time, i.e. $\overline{x^2} =2 D t$, in which $D$ is the diffusivity; and $(ii)$ Gaussian shape for tracer distribution in  space. Numerous experiments on tracer diffusion reported systems that may break one or both Brownian features. Particularly, when the linear time growth for MSD is violated, we obtain the anomalous diffusion process, namely \cite{metzler2000random}
\begin{eqnarray}
\overline{x^2} \propto D_{\alpha} t^{\alpha}, \label{eq:anomalous}
\end{eqnarray}
in which $\alpha$ anomalous diffusion exponent and $D_{\alpha}$ diffusion coefficient  $[\text{length}]^2/[\text{time}]^{\alpha}$. The  second moment is defined by  $\overline{x^2} = \int_{\Omega} p(x,t) x^2 dx$, with $p(x,t)$ the probability distribution of tracers, and $\Omega$ the spatial variable domain.    More specifically, for $\alpha=1$ we recover the usual Brownian diffusion, for  $\alpha<1$ we have the sub-diffusion, and for $\alpha>1$, the super-diffusion. 
The anomalous diffusion emerges in many theoretical and experimental investigations,  each one with their mechanisms \cite{metzler2000random,dos2019analytic,oliveira2019anomalous}. 

Anomalous diffusion may be accompanied by non-Gaussian probability density function (PDF) in space for  the tracer particles. For instance, fractional diffusion processes and nonlinear diffusion \cite{dos2019analytic,dos2019fractional,dos2018non,sene2019analytical,hristov2011approximate,bologna2000anomalous,tsallis1996anomalous}, such diffusion process breaks both Brownian features. There are also the formalisms that imply anomalous diffusion accompanied by Gaussian shape, such as the scaled Brownian motion (sBm) 
\cite{lim2002self,bodrova2019scaled,jeon2014scaled,safdari2015quantifying,tawfik2021generalized} and fractional Brownian motion (fBm) \cite{mandelbrot1968fractional,deng2009ergodic,jeon2010fractional}. Particularly, the research \cite{deng2009ergodic} on ergodicity analyse was corrected \cite{wang2020fractional} and generalised \cite{PhysRevE.104.024115}. Currently, experimental data with single-tracking particles revealed a diffusion class whose diffusion has a non-Gaussian shape for tracer displacements in space; however, the MSD increases linearly with time \cite{wang2012brownian}. Such systems have been investigated through the superstatistical approach to Brownian particles \cite{chechkin2017brownian,metzler2017gaussianity,chakraborty2020disorder,metzler2020superstatistics,slkezak2019random,lanoiselee2019non,molina2016fractional,mackala2019statistical}. In a general overview, superstatistics was proposed in the non-equilibrium statistical physics context. Nowadays, it is useful to describe many complex systems  \cite{gravanis2021blackbody,thompson2008plant,chung2021comparison,hassanabadi2021superstatistics,agahi2020truncated,agahi2020tsallis,schafer2018non,dos2021probability,dos2020mittag,DOSSANTOS2020,akilli2021wavelet,vitali2018langevin}. In the random mobility framework, there are also other formalisms that have been considered, as the ``diffusing diffusivity" model \cite{chubynsky2014diffusing,jain2016diffusing}\textcolor{red}{,} and the subordination model \cite{fogedby1994langevin,chechkin2021relation}.

The diffusion of Brownian particles in heterogeneous scenarios allows us to consider a probability density function (PDF) for diffusivities $f(D)$ to approach the problem from two different point of views \cite{metzler2020superstatistics}: {\it  heterogeneous ensemble of tracers} and {\it identical tracers in a heterogeneous medium} with each particle diffusing into their own patch. The heterogeneous environment interpretation corresponds to the superstatistics approach only when particles do not hit on the boundary of the patch. If particles explore other patches, the system undergoes a Gaussian statistic \cite{metzler2020superstatistics,dos2021random}. However, regardless of superstatistic interpretations, the system presents a non-Gaussian shape and usual diffusion.  Particularly, the superstatistics for the heterogeneous ensemble of particles governed by sBm and fBm was addressed in Ref. \cite{DOSSANTOS2021111422}. Beyond of the distributed diffusivities for particles (or patches), the particles may have their own anomalous diffusion exponent $\alpha$ (see Eq. (\ref{eq:anomalous})), such as discussed in Ref. \cite{BeckItto}. This formulations imply a statistic for anomalous exponents, i.e., a PDF $g(\alpha)$. Experiments with  single-particle tracking in complex biological media have presented heterogeneity of $\alpha$ exponent, for instance in tracer-particles diffusing in the cytoplasm of mammalian cells \cite{janczura2021identifying,sabri2020elucidating,etoc2018non}, heterogeneous population of amoeboid cells \cite{cherstvy2018non} and polydisperse vacuoles
in highly motile amoeboid cells \cite{thapa2019transient}.  
   Others problems which are not connected to biological scenarios and  exhibit distinct anomalous exponents can be found in the diffusion of trajectories in chaotic Hamiltonian systems \cite{palmero2021sub,manos2014survey}. 

In the anomalous-exponent statistics framework, our proposal here is to motivate the two most usual PDFs of anomalous exponent that define novel superstatistics, consequently showing the non-Gaussian and MSD behaviour for tracer diffusion governed by sBm. During our analysis, we will show the importance of changing the time scale to address cases where anomalous diffusion occurs for the complete ensemble. Another important point is to introduce the conditioned case between the anomalous exponent and diffusivity, this type of feature has been reported in some works in recent years  \cite{cherstvy2019non,speckner2021single,janczura2021identifying,sabri2020elucidating,etoc2018non,cherstvy2018non,thapa2019transient}. 

This article is organised in the following way: in Section (\ref{sec2}), we introduce the general setup on the statistical approach of anomalous exponent, showing what has been investigated in theoretical and experimental scenarios. Furthermore, we show how the superstatistical approach provides the necessary tools to describe a heterogeneous ensemble of tracers governed by sBm. 
In Section (\ref{sec3}), we analyse carefully the role of the anomalous exponent statistics for a mixture of anomalous tracers. 
Then, we present two truncated superstatistics for the anomalous exponent as well as how such insights may be applied to obtain different classes of the anomalous non-Gaussian diffusion processes. In Section (\ref{sec4}), we open a discussion about the importance of a generalised temporal scaling in the tracers (sBm) as well as the conditioned distribution of anomalous exponent and diffusivity. 
Finally, in Section (\ref{sec5}), we present the conclusions and how our approach contributes to random diffusivity scenarios.

\section{\label{sec2} General setup on statistics of anomalous tracers}

 Recent experiments and simulations have reported tracers diffusion with heterogeneous mobilities in which each trace presents different $\alpha$-anomalous exponents, implying a PDF~$g(\alpha)$. For instance, such features have been reported for particles diffusing in biological matter  \cite{brauchle2002single,seisenberger2001real,janczura2021identifying,sabri2020elucidating,weiss2019resampling,etoc2018non,golan2017resolving,korabel2021local}, diffusion in agitated fluids \cite{gires2020quantifying}, heterogeneous process   \cite{cherstvy2015ergodicity,xu2020heterogeneous}, and random walks with correlated noise \cite{kepten2013improved,BeckItto}. In this scenario, there are some sophisticated developments to describe the mixture of heterogeneous tracers in complex diffusion. A current way is by considering $\alpha$ anomalous diffusion exponent as a random variable, allowing us to approach systems whose anomalous exponent is randomly chosen, i.e., each tracer has its  own diffusivity and anomalous exponent such as in superstatistics interpretation  \cite{itto2014heterogeneous,BeckItto}. Investigation of dynamics with random anomalous exponent emerges also in different theoretical landscapes, such as: fractional Brownian motion (fBm) \cite{Kepten2011,Burnecki2012}, the continuous time random walkers  \cite{massignan2014nonergodic}, and the heterogeneous-diffusion process with position-dependent diffusion coefficient \cite{xu2020heterogeneous}. These systems open the way to the investigation of generalised pictures. In this work, we focus our attention on the superstatistics approach of anomalous exponent $\alpha$ and generalised diffusivity $D$ of heterogeneous tracers governed by the scaled Brownian motion, i.e., $dx/dt = \sqrt{2 \mathcal{D}(t)}\eta_t$, in which $\eta_t$ is the white noise and $\mathcal{D}(t) = D \alpha t^{\alpha-1}$ is the time-dependent diffusion coefficient. Different names for $\mathcal{D}(t)$ and $D$ can be found in literature \cite{jeon2014scaled}, for instance in the superstatistics framework, in Ref. \cite{DOSSANTOS2021111422} the authors wrote   $\mathcal{D}(t)=\mu \lambda(t)$, where $\mu$ is mobility and $\lambda(t)$ represent a general time scale. Here we call $\mathcal{D}(t)$ and $D$ by time-dependent diffusion coefficient and diffusivity, respectively. Recalling that $D$ is a generalised diffusion coefficient due to the physical dimensions, see Eq. (\ref{eq:anomalous}).    

 The {\it ensemble} of heterogeneous particles (or tracers) will imply a distribution of anomalous exponents and diffusivities that define the two-variable superstatistical approach  \cite{BeckItto}. Here, we consider that all tracers are governed by sBm, and each tracer with its anomalous exponent and diffusivity. Hence, the overall PDF of the tracers is given by
\begin{eqnarray}
p(x,t)=\int_{\Omega_{\alpha}}\int_{\Omega_{D}} \pi(\alpha,D) \mathcal{G}_{\alpha}(x,t|D) dD d\alpha, \label{eq:overallPDF}
\end{eqnarray}
in which $\alpha$ is the anomalous diffusion exponent, and $D$, the diffusivity, with $\Omega_{\alpha}$ and $\Omega_{D}$ being the domains, respectively. The PDF $\pi(\alpha, D)$ is the joint distribution that defines a  superstatistic, such PDF contains all information about $\alpha$ and $D$  variables. Moreover,  $\mathcal{G}_{\alpha}(x,t|D)$ is the PDF of tracers with indexed $\alpha$ and $D$ parameters, which in the framework of sBm, has a Gaussian shape. Indeed, we are superposing many different subsets of anomalous tracers each one with an indexed $\alpha$ exponent. Thereby, the {\it random diffusivity} term is related to the set of generalised diffusion coefficients $\{ D\}$ that implies heterogeneity of the tracers. 
This superstatistical approach provides an efficient way to address experimental data tracking associated with a mixture of tracers. There are considerations associated to different approximations on PDF of parameters, i.e. $\pi(\alpha,D)$. For instance,  $\pi(\alpha,D)=g(\alpha)f(D)$ corresponds to the case in which $\alpha$ and $D$ are  uncorrelated parameters, considering $g(\alpha)$ and $f(D)$ as the PDFs of anomalous exponents and diffusivities, respectively. Another complex scenario occurs when the anomalous diffusion exponent is conditioned to diffusivity, i.e. $\pi(\alpha,D)=g(\alpha|D)f(D)$, as it was approached in superstatistical treatment on two variables. Current applications about it were observed in data of protein diffusing within a bacteria \cite{BeckItto}. There are also experimental indicators of correlations between diffusivities and anomalous exponents in the spreading dynamics of amoeboid cells  \cite{cherstvy2018non,thapa2019transient}.  In such situations the authors found an exponential adjustment which connects both quantities, e.g., $D_{\alpha}\sim e^{ -c_1 \alpha +c_2 }$, for more details see Ref. \cite{cherstvy2018non}. The relation between $\alpha$ and $D$ have been observed in others recent researches \cite{cherstvy2019non,speckner2021single}.

\begin{figure}[h]
\centering
\includegraphics[width=0.8\textwidth]{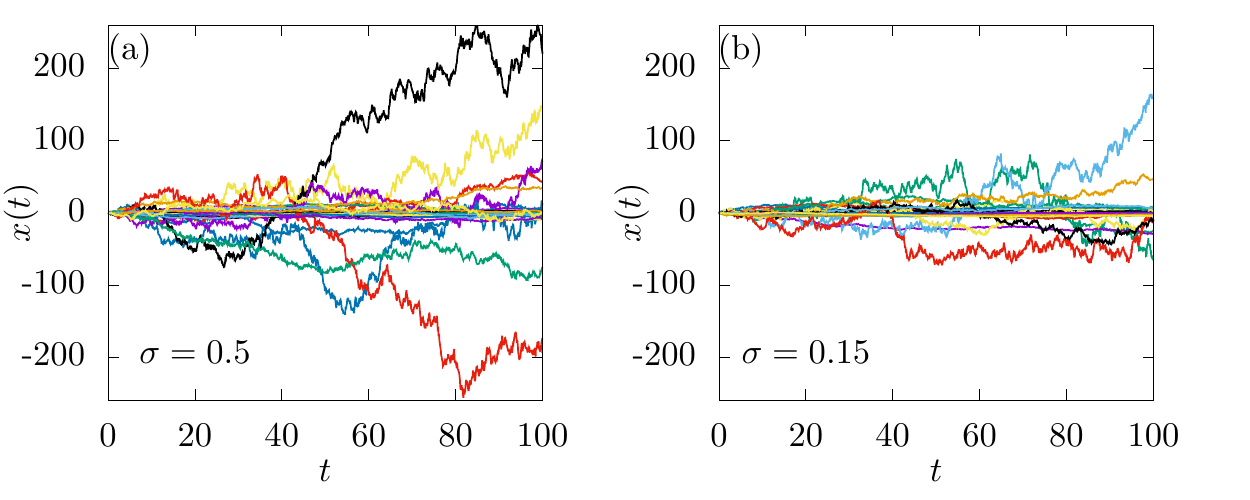}
\caption{\small{Figures show scaled Brownian tracers, each one with their own $\alpha$-anomalous exponent randomly chosen from the truncated Gaussian distribution  (\ref{gaussian}), with $\alpha \in (0,2]$ and $\alpha_p=1$. Fig. (a) shows trajectories for $\sigma=0.5$ exponent. Fig. (b) shows trajectories for $\sigma=0.15$ exponent.  
} }
\label{Fig1}
\end{figure}

The overall PDF of tracers depends on interpretation, heterogeneous anomalous tracers or identical tracer-particles moving in their own patch with indexed mobility. 
  All tracers associated to fixed $(\alpha, D)$ values that imply a particular average $\overline{x^2_{\alpha}}$. Therefore, by mixing the MSD profiles with different $\alpha$ and $D$ the overall MSD in the superstatistics sense is given by
\begin{eqnarray}
\left\langle \overline{x^2 }  \right\rangle = \int_{\Omega_{\alpha}}\int_{\Omega_{D}}  \pi(\alpha,D) \overline{x^2_{\alpha}} dD d\alpha, \label{msdMix}
\end{eqnarray}
in which $\left\langle \overline{x^2 }  \right\rangle =\left\langle \overline{x^2_{\alpha} }  \right\rangle$. Keeping in mind that $\overline{x^2_{\alpha}}$ emerges from an arbitrary generalised dynamic that implies power-law behaviour for time, i.e., $\overline{x^2_{\alpha}}\sim D t^{\alpha}$, or similar ones. Figure. (\ref{Fig1}) shows $30$ trajectories  generated by scaled Brownian process ($dx/dt = \sqrt{2\alpha t^{\alpha-1}}\eta_t$, in which $\eta_t$ is the white noise) for $dt=0.1$, step number $N=1000$ and truncated Gaussian PDF of $\alpha$ values with two different parameters. The truncated Gaussian of anomalous exponents is defined by $g_n(\alpha)=\mathcal{N}^{-1} \exp(-(\alpha-\alpha_p)^2/2\sigma^2 ) $ with $\alpha \in (0,2]$, in which  $\mathcal{N}$ is the normalisation constant and $\sigma$ controls the shape of PDF of the anomalous exponents. In Figure (\ref{Fig1}-a), we consider that $\alpha$ is randomly chosen from a truncated Gaussian distribution with $\sigma=0.5$, see the Eq. (\ref{gaussian}). Fig. (\ref{Fig1}-b) considers a similar situation but with $\sigma=0.15$. Such graphical representations reveal to us a rich scenario for the mixed scaled Brownian motion, the truncated Gaussian PDF, as well as other anomalous exponent statistics, will be carefully investigated in the next sections. The sBm may admit yet a general time-dependent diffusion coefficient  \cite{bodrova2016underdamped} which allow us to introduce a general temporal scale $\psi(t)$ thought the relation $ \mathcal{D}(t) = D \frac{d\ }{dt}\psi(t) \propto \frac{d\ }{dt}\overline{x^2_{\alpha}}$, the role of this general scale will be studied in this work. It is noteworthy that scaled Brownian motion  establishes an anomalous diffusion process based on the overdamped Langevin equation \cite{molina2016fractional,mackala2019statistical,bodrova2016underdamped}.

\section{Heterogeneous anomalous exponent for scaled  diffusion   }
 \label{sec3} 
Under the general setup of $\alpha$-statistics previously  discussed, we have an open way to new theoretical diffusion approaches. In this section, we analyse the $\alpha$-statistics effects on a tracer diffusion in a complex environment. To do this, we disregard the correlation effect which might exist experimentally between $\alpha$ and $D$, so that random diffusivity does not overshadow the effect of the fluctuations of the anomalous exponent on sBm. Consequently, we have the approximation for the $\pi(\alpha,D)$ statistic of  parameters
\begin{eqnarray}
\pi(\alpha,D) = g(\alpha) f(D),
\end{eqnarray}
such PDF describes the scenarios in which the anomalous diffusion exponent is not conditioned to diffusivity. 

\begin{figure}[h]
\centering
\includegraphics[width=0.8\textwidth]{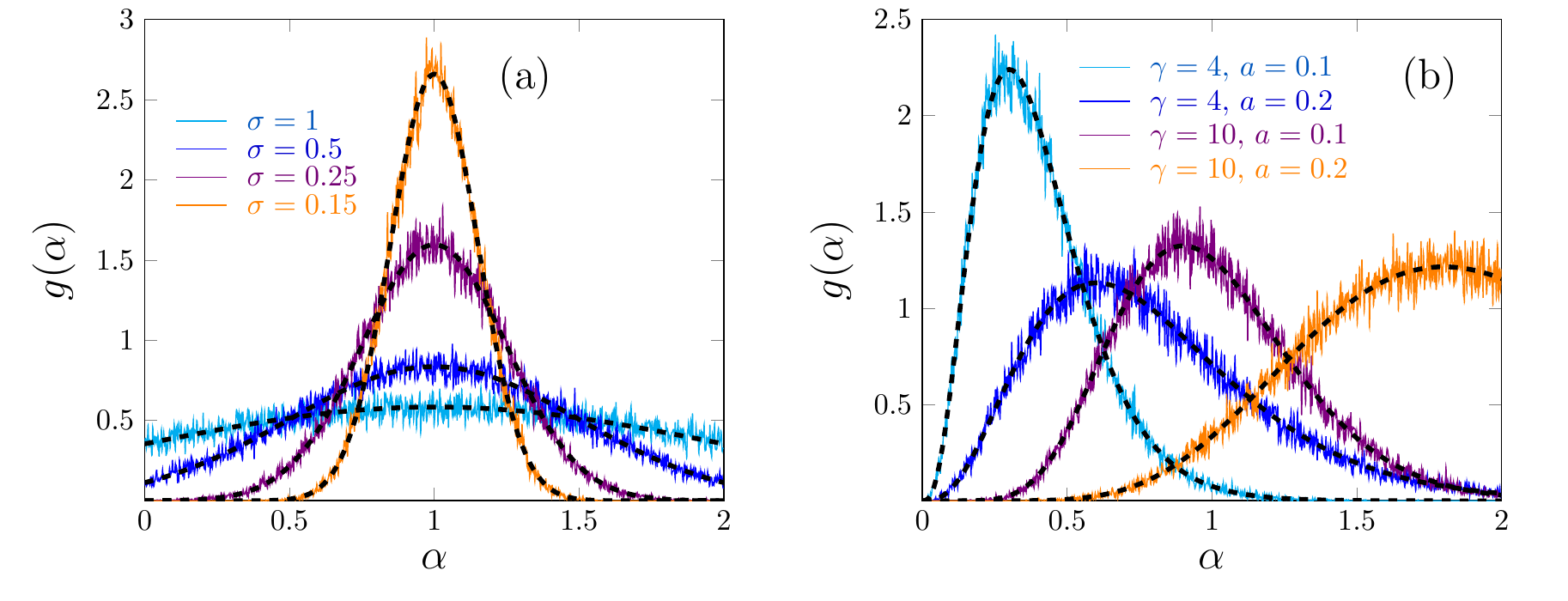}
\caption{\small Figures present PDF of the anomalous diffusion exponent for (a) truncated Gaussian statistic (with $\alpha_p=1$) and (b) truncated $\chi^2$-Gamma statistic. The dashed lines represent the analytical approach (see Eqs. (\ref{gaussian}) and (\ref{gamma})). The coloured curves were obtained from simulations.}
\label{Fig2}
\end{figure}

The PDF $g(\alpha)$ of the anomalous exponent was investigated through the overall MSD  in Eq. (\ref{msdMix}) with $f(D)=\delta(D-D_0)$. For different statistics of $\alpha$-exponent, see Ref. \cite{li2021statistics}. A relevant feature here is that for $\alpha$ contained in an infinite domain, the limit $\lim_{\alpha\to \infty }g(\alpha)$ must converge to zero, so that the overall MSD $\left\langle \overline{x^2} \right\rangle$ cannot diverge from any time value. To avoid the divergence effects at MSD calculus, we suggest two truncated $\alpha$ statistics. The first one, is the truncated-normal statistics (or truncated-Gaussian statistics) of the anomalous diffusion exponent, defined as follows
\begin{eqnarray}
g_{n}(\alpha) =
  \begin{cases}
    \displaystyle \mathcal{N}^{-1} \exp\left(-\frac{(\alpha-\alpha_p)^2}{2\sigma^2} \right)  &  \quad \text{if }  0 \leq \alpha \leq \alpha_{\text{max}} \\
    0  & \quad  \text{ otherwise}
  \end{cases}
  \label{gaussian}
\end{eqnarray} 
in which $\mathcal{N}=\sqrt{\frac{\pi }{2}} \sigma  \left(\text{erf}(\alpha_p (\sqrt{2} \sigma)^{-1})+\text{erf}((\alpha_{\text{max}}-\alpha_p)(\sqrt{2} \sigma)^{-1})\right)$ is the normalisation factor, $\sigma$ controls the distribution width, and $\alpha_p$ defines where the peak of the distribution is. The Fig. (\ref{Fig2}-a) exemplifies the qualitative features of $g(\alpha)$ distribution. Thereby, for lower values of $\sigma$-variance the PDF (\ref{gaussian}) leads to a concentrated delta distribution, i.e.   $\lim_{\sigma\to 0}g(\alpha)\sim \delta(\alpha-\alpha_p)$ as well as the standard scaled Brownian diffusion with $\alpha_p$. The Gaussian statistic in Eq. (\ref{gaussian}) for anomalous exponents was approached in experiments on single tracking particles diffusing within an environment with micro pillars \cite{chakraborty2020disorder}, particles diffusing in a heterogeneous environment \cite{xu2020heterogeneous,cherstvy2015ergodicity}, and biological specimen diffusing in soft-matter \cite{weiss2019resampling}.

Now, we consider the second statistic, which is the truncated-$\chi^2$ gamma distribution, given by
\begin{eqnarray}
g_{\chi}(\alpha) =  \begin{cases}
    \displaystyle \mathcal{N}^{-1}\alpha^{\gamma-1} \exp \left(-\frac{\alpha}{a} \right)   & \quad \text{if }  0 \leq \alpha \leq \alpha_{\text{max}} \\
    0  & \quad  \text{ otherwise}
  \end{cases}
  \label{gamma}
\end{eqnarray} 
in which $\mathcal{N}= a^{\gamma} \left(\Gamma (\gamma )-\Gamma \left(\gamma ,\alpha_{\text{max}} a^{-1}\right)\right)$ is the normalisation condition, $\gamma$ and $a$ are free parameters that control the shape of distribution, see Fig. (\ref{Fig2}). The $\chi^2$-Gamma PDF in superstatistics was applied in protein movement in biological matter \cite{BeckItto}. There are other problems where $\chi^2$ PDF for the anomalous exponent seems present in experimental or numerical data of tracking particles, such as telomeres diffusing in the nucleus of mammalian cells \cite{stadler2017non}, heterogeneous diffusion states in the cytoplasm \cite{janczura2021identifying} and heterogeneous diffusion processes \cite{cherstvy2013anomalous}.

Both $\alpha$-statistics have truncation, which is crucial to ensure that the overall MSD behaviour does not go to infinity for a fixed time. Mathematically, the integration (\ref{msdMix}) on $\alpha$ values to a fixed time cannot diverge; therefore,  truncation ensures the MSD convergence. From a  physical point of view, the truncation condition guarantees us that there are not individual tracers with unrealistic values of $\alpha$. Thereby, there are individual tracers with anomalous diffusion exponents that are sub or superdiffusive, but never with $\alpha>\alpha_{\text{max}}$, in which $\alpha_{\text{max}}$ was imposed. In most experimental situations, tracers with $\alpha>3$ represent a  negligible number; therefore, the truncation works well under this condition.

The model considers heterogeneous tracers with their own indexed $\alpha$ exponent, i.e., $\alpha_i$ exponent that remains fixed the whole time. Individually, each tracer is generated by the  following scaled-stochastic process
\begin{eqnarray}
\frac{dx_i}{dt} = \sqrt{2D_i \alpha_i t^{\alpha_i-1}} \eta(t), \label{Xmodel1}
\end{eqnarray}
in which $t>0$. The $\alpha_i$ and $D_i$ are different for each $i$-indexed tracer. The parameter $D_i$  is the generalised diffusion coefficient with dimension $[\text{length}]^2 / [\text{time}]^{\alpha}$. The tracer’s displacement on an individual level obeys a Gaussian diffusion process, given by
\begin{eqnarray}
\mathcal{G}_{\alpha_i}(x,t|D_i)=  \frac{1}{\sqrt{4\pi D_i t^{\alpha_i}}} \exp \left( - \frac{x^2}{4 D_i t^{\alpha_i} } \right).
\end{eqnarray}
This distribution has the same mathematical form of the PDF associated with fractional Brownian motion \cite{PhysRevE.104.024115}.
To analyse the PDF of tracers in space only with anomalous exponent effects, we approximate  $f(D) \approx  \delta(D-D_0)$ to remove the diffusivity effects. Thereby, we consider two $\alpha$ distributions of the anomalous diffusion exponent that are not conditioned to diffusivity and have a finite domain. 
The mix of all tracers is defined by the overlapping of the heterogeneous sBm, as shown in Eq. (\ref{eq:overallPDF}). 
 In this situation, we reduced our problem to a one-variable superstatistics, which implies 
\begin{eqnarray}
P(x,t) \simeq \int_0^{\alpha_{max}}   \mathcal{G}_{\alpha}(x,t|D_0) g(\alpha)  d\alpha. \label{alphasuperstatistics}
\end{eqnarray}
This PDF of tracers allow us to define a \textit{pure} superstatistics on the anomalous exponent, i.e., $f(D) \approx  \delta(D-D_0)$. The MSD in Eq. (\ref{msdMix}) for this case is given by
\begin{eqnarray}
\left\langle \overline{x^2} \right \rangle =   2 D_0 \int_0^{\alpha_{max}} t^{\alpha} g(\alpha)  d\alpha. \label{msd:alphasuperstatistics}
\end{eqnarray}
recalling that we adopt the notation $\langle x^2 \rangle = \left\langle \overline{x^2} \right \rangle$ for the figures. 
For $g(\alpha)$ different from Dirac-delta distribution, we have non-Gaussian diffusion.

\begin{figure}[htp]
\centering
\includegraphics[width=0.8\textwidth]{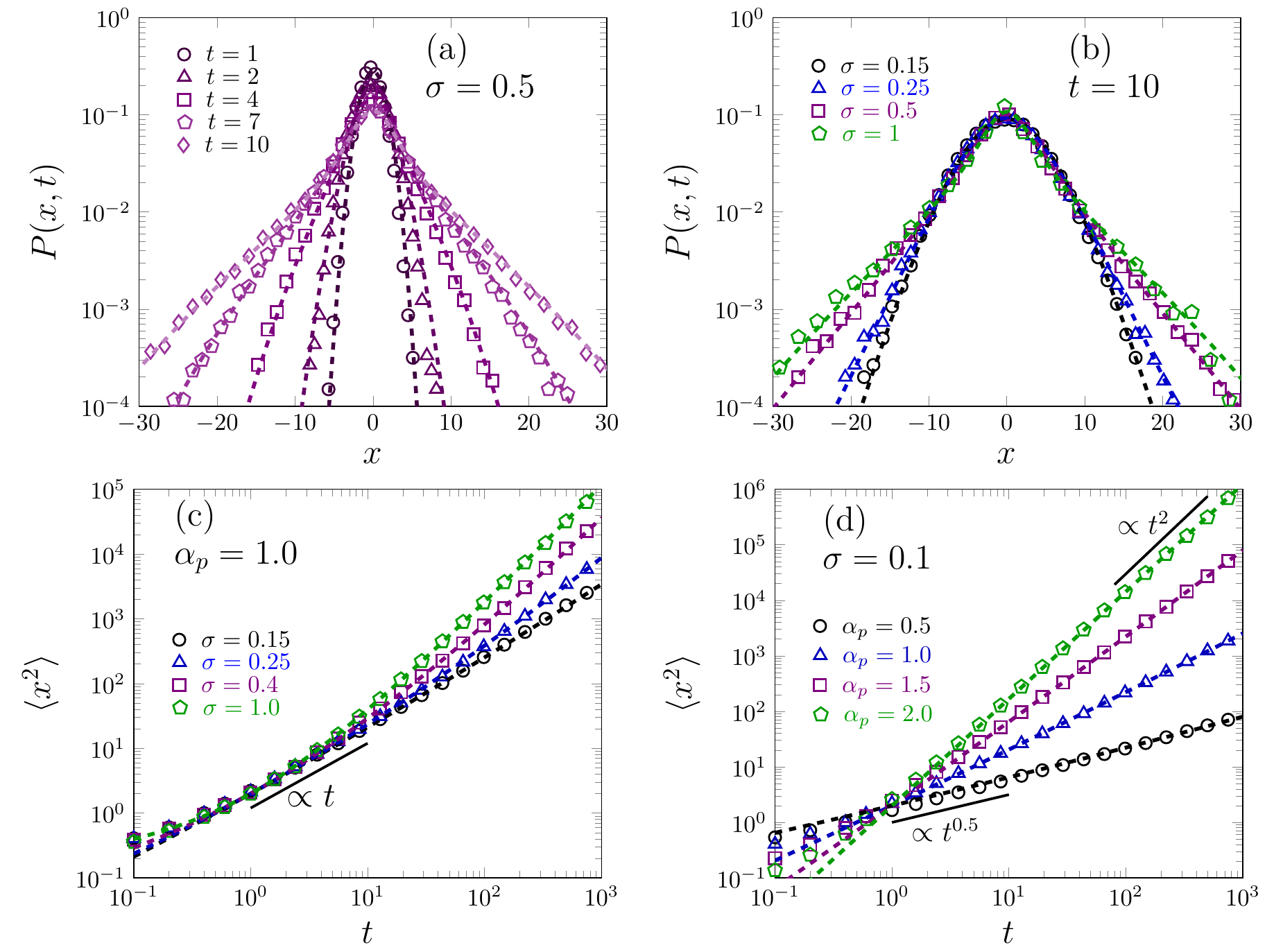}
\caption{Figures show overall PDFs of tracers in space and MSD behaviour in time. All figures consider truncated Gaussian statistics for the anomalous exponent $\alpha\in (0,2]$. The dashed curves represent the analytical approach that were obtained through of Eqs. (\ref{alphasuperstatistics}) and (\ref{msd:alphasuperstatistics}) for PDF and MSD, respectively. The symbolic curves represent the simulation results obtained by the scaled Brownian process, see Eq. (\ref{Xmodel1}). Fig. (a) shows the PDF of tracers for $\alpha_p=1.0$ and different times. Fig. (b) shows the PDF of tracers for $\alpha_p=1.0$ and different $\sigma$ values. Figs. (c) and (d) show the overall MSD of tracers in time for different $\sigma$ and $\alpha_p$ values, respectively.}
\label{Fig3}
\end{figure}

\begin{figure}[h!]
\centering
\includegraphics[width=0.8\textwidth]{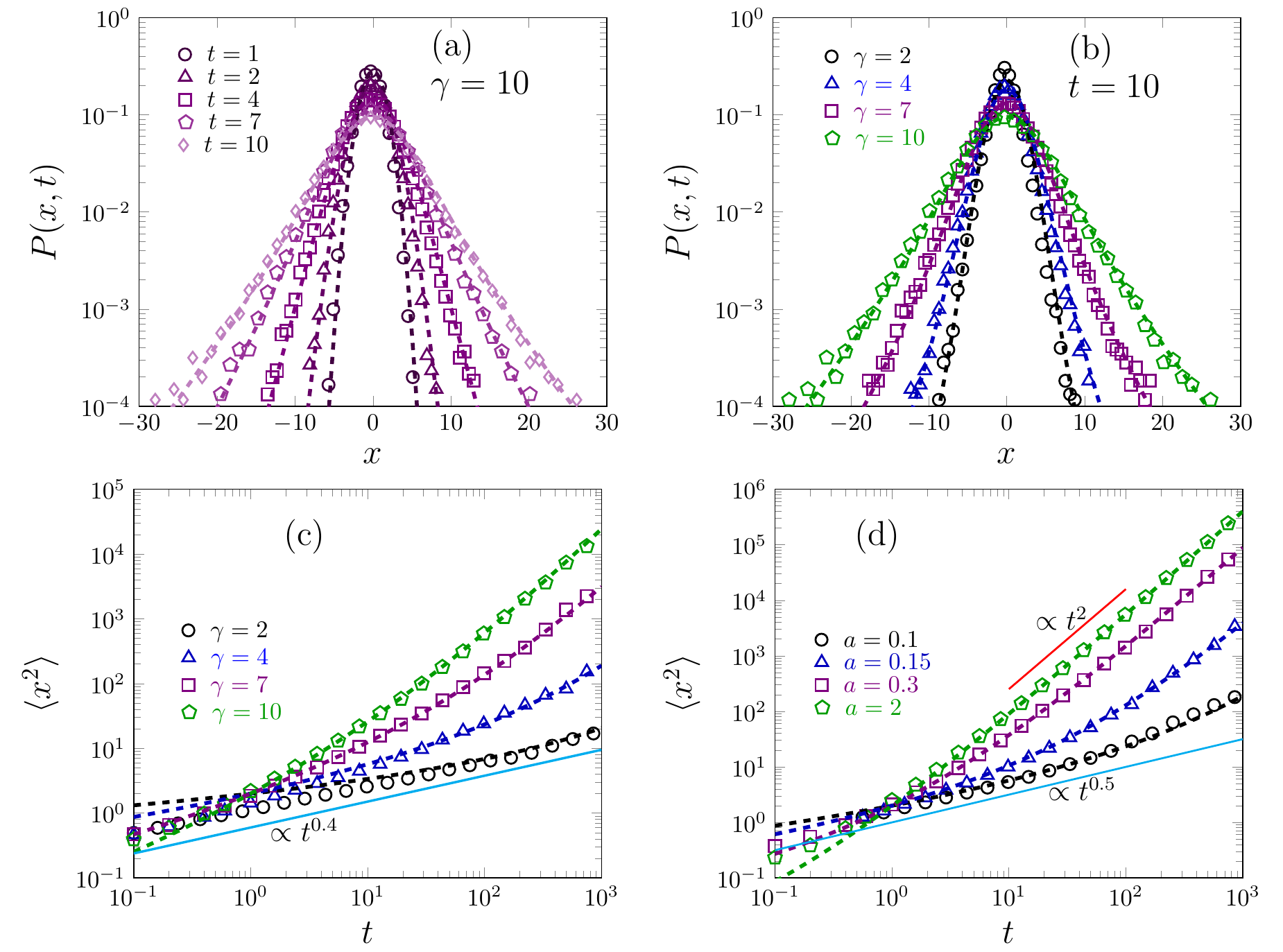}
\caption{Figures show overall PDFs of tracers in space and MSD behaviour in time. All figures consider truncated $\chi^2$-Gamma statistics for the anomalous exponent $\alpha \in (0,2]$. The dashed curves represent the analytical approach that were obtained through of Eqs. (\ref{alphasuperstatistics}) and (\ref{msd:alphasuperstatistics}) for PDF and MSD, respectively. The symbolic curves represent the simulation results obtained by the scaled Brownian process, see Eq. (\ref{Xmodel1}). Fig. (a) shows the PDF of tracers  for $a=0.1$ and different times. Fig. (b) shows the PDF of tracers for $a=0.1$ and different $\gamma$ values. Fig. (c) shows the MSD for $a=0.1$ and different $\gamma$ values. Fig. (d) shows the MSD for $\gamma=10$ and different $a$ values. }
\label{Fig4}
\end{figure}

Now, we may check the match between analytical and simulation approaches. In order to do so, we consider two PDFs of the anomalous exponent introduced in Eqs. (\ref{gaussian}) and (\ref{gamma}), and $10^4$ tracers generated by Eq. (\ref{Xmodel1}). Consequently, figures (\ref{Fig3}-a) and (\ref{Fig3}-b) present the PDF of tracers with truncated Gaussian statistic for the  anomalous exponents; the figures consider different times and  parameters. Moreover, Figs. (\ref{Fig3}-c) and (\ref{Fig3}-d) show the influence of $\sigma$ and $\alpha_p$ values on MSD. 
Figure (\ref{Fig4}) repeats the approach to the truncated $\chi^2$-Gamma distribution of anomalous exponent. Figs. (\ref{Fig4}-a) and (\ref{Fig4}-b) show the PDF of displacements in space. Figs. (\ref{Fig4}-c) and (\ref{Fig4}-d) show the influence of $\gamma$ and $a$ parameters on MSD, respectively. Both superstatistics present a perfect agreement between analytical and simulation results. An important feature here is that through the straight lines in cyan colour in Figs. (\ref{Fig4}-c) and (\ref{Fig4}-d), we notice that the MSD curves do not follow a perfect power-law behaviour in time, i.e. $\langle x^2 \rangle \nsim t^{q}$. Particularly, for $a=2$ (green curve), as shown Fig. (\ref{Fig4}-d), the MSD curve for long times is approximated to  a power-law time behaviour due to the bigger concentration of anomalous exponent values around $\alpha\sim 2$. However, for the  Gaussian PDF of the  anomalous exponent, the overall MSD presents an exact power-law  behaviour in time when $\sigma \to 0$.

\newpage

\section{Other temporal scaling into the SBM  and two-variable superstatistics} 
\label{sec4}

A heterogeneous ensemble of tracers with statistics over $\alpha$-anomalous exponent and diffusivity $D$ was considered in the sBm context in the previous section. Such systems with heterogeneity only in diffusivity were investigated to different landscapes \cite{molina2016fractional,vitali2018langevin,dos2021random,DOSSANTOS2020}. Motivated by cases investigated in the previous section, we introduce two new features in this section: firstly, a general temporal scaling; secondly, we consider the entanglement between the $\alpha$ and $D$ parameters to emulate correlated situations.

To introduce the idea of general temporal scaling in superstatistics, let us introduce a motivation connected to our problem. The MSD for the previous model increases in  time according with Eq. (\ref{msdMix}). For simplified cases of $\alpha$-superstatistics, the MSD may be calculated analytically \cite{li2021statistics}. For instance, we consider an uncoupled $\pi(\alpha,D)=f(D)g(\alpha)$  distribution with $f(D)=\delta(D-D_0)$ and $g(\alpha)$ being a uniform distribution with domain $\alpha \in [\alpha_{\text{min}},\alpha_{\text{max}}]$, applying this statistics in Eq. (\ref{msdMix}) and considering a long time limit, we obtain
\begin{eqnarray}
\lim_{t \to \infty} \left \langle \overline{x^2} \right\rangle &=& \lim_{t \to \infty} 2D_0\int^{\alpha_{\text{max}}}_{\alpha_{\text{min}}} t^{\alpha}d\alpha \nonumber   \\ 
& \simeq &  2D_0 t^{\alpha_{\text{max}}}(\log(t))^{-1}. \label{eq:logfactor}
\end{eqnarray}
This limit shows that there is a logarithmic factor that avoids a pure power law growth \cite{li2021statistics}. If we consider a  Poisson-like form of fluctuations for the $\alpha$-anomalous exponent \cite{itto2014heterogeneous,itto2012heterogeneous}, i.e. $g(\alpha)\propto\exp(\lambda\alpha)$ for $0\leq\alpha\leq 1$, the  superstatistics of MSD imply $\left\langle \overline{x^2}\right\rangle \propto t /\log(t)$ for long times.  A  mixing of fractional-time continuous random walk was considered in Ref. \cite{itto2014heterogeneous}. 
Then, we have a logarithmic factor in Eq (\ref{eq:logfactor}) that is implicitly present in numerical simulation cases approached in the previous section, see Figs. (\ref{Fig3}) and (\ref{Fig4}). Thereby, it is not so easy to obtain a power-law time behaviour for superstatistics of anomalous exponents by mixing power-law  profiles, i.e. $\int_{\Omega_{\alpha}}t^{\alpha}g(\alpha)\sim t^{\alpha_{\text{eff}}}$ with $\alpha_{\text{eff}}>0$. However, most of the experimental investigations present a MSD behaviour that increases as a power-law function of time; this lead us to consider an appropriate time scaling.

 In the previous section, a power-law time temporal scale was considered for the time-dependent diffusion coefficient. However, different profiles for $\mathcal{D}(t)$ can emerge due to intrinsic or extrinsic factors,  related to memory and aging effects or due to changes in the spatial structure of the medium \cite{sen2004time,latour1994time,safdari2015aging,cherstvy2021anomalous}. Then, other functions of the diffusion coefficient may configure different classes of temporal scales.  Thus, we consider a general function for temporal scaling  $\psi_{\alpha}(t)$ \cite{DOSSANTOS2021111422,lim2002self,cherstvy2021anomalous,DOSSANTOS2020}, in this perspective the general temporal scale embraces the power-law time scale as a particular case. Moreover, different profiles of temporal scales beyond of power-law time function, e.g., logarithmic and exponential profiles, have been motivated by the biophysical application of the diffusion models with time-dependent diffusion coefficient, namely, the diffusion of water molecules in brain tissues \cite{cherstvy2021anomalous}. The overall MSD of the tracer is given by
\begin{eqnarray}
\left\langle \overline{x^2} \right \rangle = \int_{\Omega_{D}} \int_{\Omega_{\alpha}} 2 D \psi_{\alpha}(t) \pi(\alpha,D) d\alpha dD, \label{eq:overallMSD2}
\end{eqnarray}
recalling that we adopt the notation $\langle x^2 \rangle = \left\langle \overline{x^2} \right \rangle$ for the figures. The time-dependent diffusivity is defined by  $\mathcal{D}(t)= \psi_{\alpha}'(t)D$, see Ref. \cite{DOSSANTOS2021111422}.

In this framework, we can choose the temporal scale $\psi(t)$ to change the effective diffusivity, allowing us to control the short and long times limits of overall MSD. In  Ref. \cite{cherstvy2021anomalous}, the authors explore two different profiles of time-dependent diffusivity following logarithmic or exponential forms.  We consider a multiplicative logarithmic factor in scaling-law of sBm to obtain an  exact power-law behaviour of MSD for long times. Thereby, we propose the following explicit form for temporal scale $
\psi_{\alpha}(t) = \left(1+t/\tau\right)^{\alpha}\log\left(1+t/\tau \right)$,
in which $\tau$ is a temporal constant $[\text{time}]$. We  have the following limits: for short times, ($t \ll \tau$) $\psi_{\alpha}(t) \sim t$; for long times, ($t \gg \tau$) $\psi_{\alpha}(t) \sim  t^{\alpha}\log t$. It is noteworthy that for short times the effect of the $\alpha$ exponent is negligible. Particularly, our first choice for the  temporal scale have special particular cases, such as $\psi(t) \propto t \log[t/\tau]$ that was approached in Ref. \cite{cherstvy2021anomalous}, and the time scale $\psi(t) \propto \log(1+ t/\tau)$ investigated in Ref. \cite{bodrova2016underdamped} motivated by ultraslow diffusion processes. However, we did this chosen because the logarithmic term is important to cancel the term $(\log(t/\tau))^{-1}$ that will appear in the overall MSD, such as appeared in the example on Eq. (\ref{eq:logfactor}). Thereby, the superstatistics over the $\alpha$ variable in Eq. (\ref{eq:overallMSD2}) imply two limits which may be determined separately. For short times, $\left\langle \overline{x^2} \right\rangle \simeq 2 D_{\text{eff}} t$, in which the effective diffusivity is given by 
\begin{eqnarray}
D_{\text{eff}} = \int_{\Omega_D}\int_{\Omega_{\alpha}}D\pi(\alpha,D)d\alpha dD. \label{eq:effectve}
\end{eqnarray}
For long time limits, we have as a general result $\left\langle \overline{x^2} \right\rangle \propto t^{\alpha_{\text{eff}}}$, in which $\alpha_{\text{eff}}$ depends on the $\alpha$ statistic. In another way, we can consider a different temporal scale $\psi_{\alpha}(t)$, which presents a usual diffusion over long times. To obtain a transition between two different diffusive process that finishes in usual behaviour, we consider $\psi_{\alpha}(t)= te^{-\tau/t} + t^{\alpha}e^{-t/\tau}$, that implies a crossover from  anomalous to usual diffusion, e.g. $\left\langle \overline{x^2} \right\rangle \simeq 2 D_{\text{eff}} t$ for long times; $D_{\text{eff}}$ is defined in Eq. (\ref{eq:effectve}). This kind of crossover out of the superstatistical context was considered in Ref. \cite{safdari2015aging,dos2021random}. Both temporal scales considered here are in Fig. (\ref{Fig7}) for a general situation that admits a coupling between $\alpha$ and $D$. Now we will explore in more detail the coupled situations between $\alpha$ and $D$.

To understand the effects of $\alpha$ on $D$ or vice versa, we consider a coupled situation for the PDF of the anomalous exponent, introducing an $\alpha$-PDF conditioned to diffusivity values, namely
\begin{eqnarray}
\pi(\alpha,D) = \Tilde{g}(\alpha|D)f(D),
\end{eqnarray}
considering $f(D)=\int_{\Omega_{\alpha}}\pi(\alpha,D)d\alpha $, and $ \int_{\Omega_{\alpha}} \Tilde{g}(\alpha|D) d\alpha =1$.  This setup to conditioned distribution emulates a correlation  between fluctuating parameters, such as discussed in two-variable superstatistics treatment \cite{BeckItto}. The distribution of anomalous exponents may be obtained by the following integration $ g(\alpha) = \int_{\Omega_{\alpha}} \pi(\alpha,D) dD$.  Here, we also consider the conditioned $\Tilde{g}(\alpha|D)$ PDF as a Gaussian superstatistic for the  anomalous exponent, as follows
\begin{eqnarray}
\pi(\alpha, D) =  \mathcal{N(D)}^{-1} \exp\left(-\frac{(\alpha-\alpha_p)^2}{2\sigma_D^2} \right) f(D), \label{eq:GassianDalpha}
\end{eqnarray}
in which $\mathcal{N}(D)=\int_{\Omega}\exp\left(-(\alpha-\alpha_p)^2(2\sigma_D^2)^{-1} \right)d\alpha$. The $\sigma_D$  contain the correlations between $\alpha$ and $D$. 
For $\sigma=\text{const}$ regardless of $D$ we recover the standard cases approached previously. Thereby,  $\sigma(D)$  explicitly depends on the diffusivity, because the $\pi$-superstatistics in Eq. (\ref{eq:GassianDalpha}) may emulate systems whose correlations between the anomalous exponent and diffusivity are unknown, but exist. 

The PDF of diffusivities $f(D)$ may assume numerous shapes as reported in \cite{chechkin2017brownian}. The main goal now is to introduce a PDF $f(D)$ to exemplify how the distribution of diffusivity impacts the distribution of anomalous exponent. To do that, we chose the most famous distribution, which is the exponential PDF, namely
\begin{eqnarray}
f(D) & = & \frac{1}{D_0 }\exp\left( - \frac{D}{D_0}  \right). \label{eq:exp}
\end{eqnarray}
The superstatistical approach to Gaussian packets with exponential PDF for diffusivities implies a Laplace diffusion. Therefore, the exponential diffusivity distribution was reported in a large number of  investigations  \cite{wang2009anomalous,wang2012brownian,metzler2017gaussianity,chubynsky2014diffusing}. Sometimes the Laplace diffusion occurs only when the observable time is shorter than the correlation time of the system. Here, we extend such analysis to a general situation in which the anomalous exponent has a distribution conditioned to diffusivity values.

\begin{figure}[h!]
\centering
\includegraphics[width=0.8\textwidth]{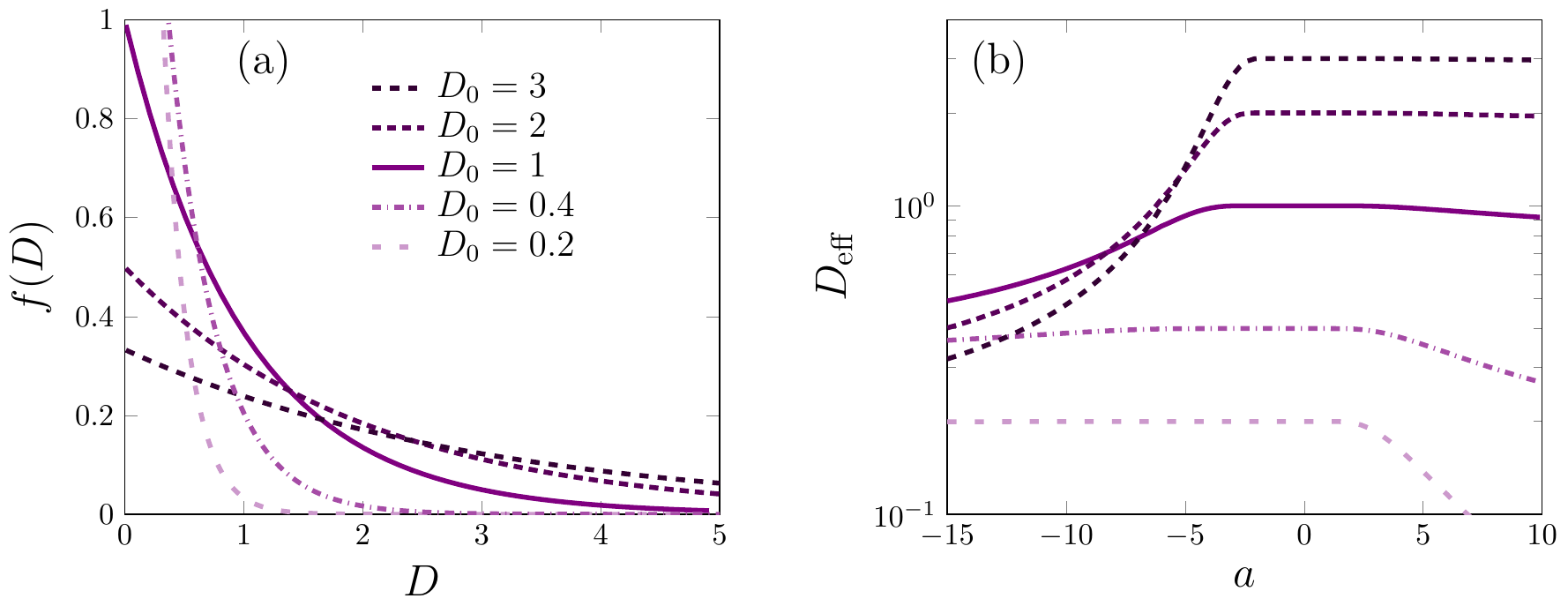}
\caption{Fig. (a) shows PDF of diffusivities for different  $D_0$ values. Fig. (b) shows the influence of $D_0$ values on effective diffusivity, resulted by Eq. (\ref{eq:GassianDalpha}) into (\ref{eq:effectve}). Both figures consider  $\sigma_0/d_0^{a}=1$, $\alpha_p=1$ and  $\alpha \in (0,2]$. }
\label{Fig5}
\end{figure}

To continue our discussion, we consider Eq. (\ref{eq:exp}) to  describe the heterogeneous  diffusivities of the tracer, and (\ref{eq:GassianDalpha}) for the PDF of the anomalous exponents with 
\begin{eqnarray}
\sigma_D=\sigma_0 (D/d_0)^{a}, \label{eq:coupled}
\end{eqnarray}
in which $\sigma_D$ controls the shape of $g(\alpha)$ PDF, $\sigma_0$ and $d_0$ are constant parameters, and $a\in \Re$ is the coupling parameter. For $a=0$, we recover the uncoupled case. Fig. (\ref{Fig5}-a) show PDF $f(D)$ for different $D_0$ values and Fig. (\ref{Fig5}-b) shows the influence of the $D_0$ values in $g(\alpha) = \int_{\Omega_D}\pi(\alpha,D) dD$. This figure exemplifies the $D_0$ influence on the effective diffusivity (\ref{eq:effectve}) for different coupling values in anomalous exponent distribution. For the lower values of $a$ and $D$, we have constant behaviour for $D_{\text{eff}}$, and for higher values for $a$ and $D$, the $D_{\text{eff}}$ presents an upper bound. However, we obtain different results for $a<0$ and large $D_0$ values; the effective diffusivity becomes smaller. Another interesting effect happens when $a$ is large and $D_0$ is small, because the diffusivity of particles chosen from $f(D)$ are always small with large $a$ values, which implies that $\alpha$-PDF becomes more and more concentrated around $\alpha\sim \alpha_p$, and in such circumstances $D_{\text{eff}}$  gets smaller.

\begin{figure}[h!]
\centering
\includegraphics[width=0.8\textwidth]{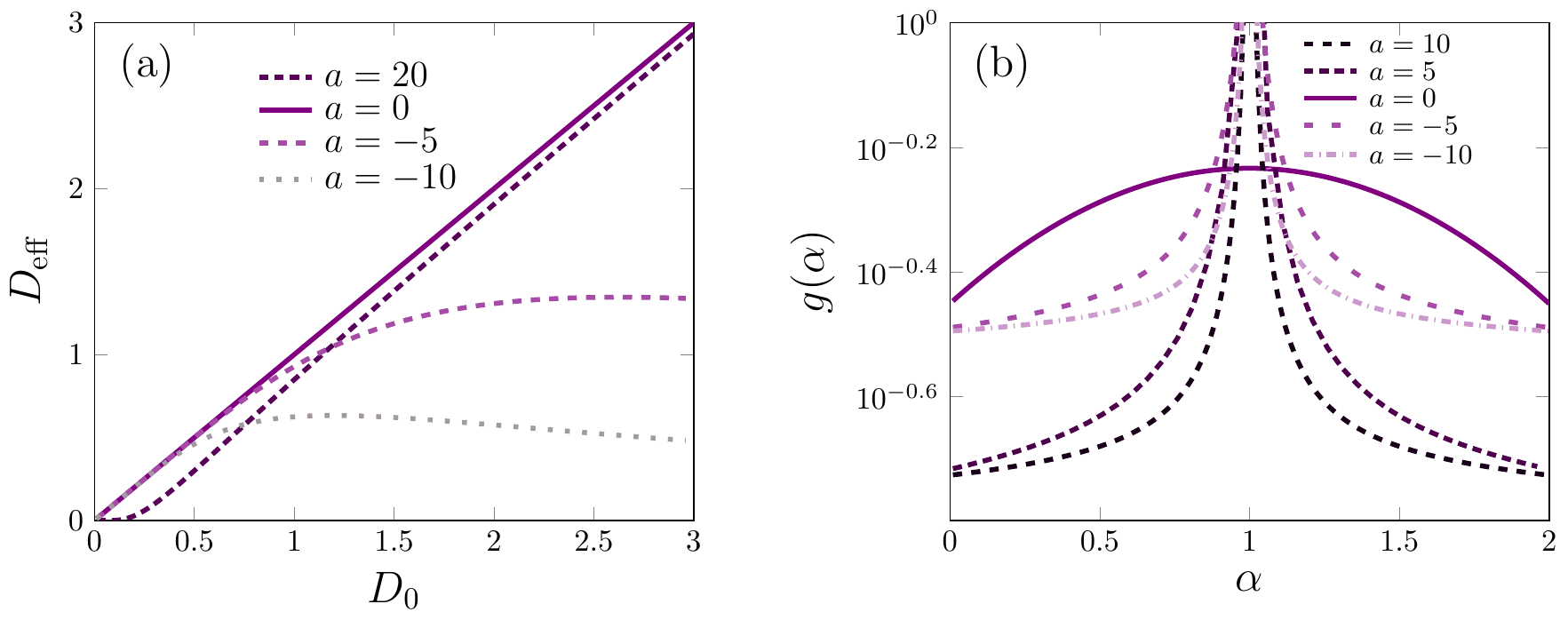}
\caption{Fig. (a) shows effective diffusivity for different  coupling parameters $a$. Fig. (b) shows the influence of $a$ values  on PDF of anomalous exponents, obtained by integration $g(\alpha)=\int_{\Omega_{\alpha}}\pi(\alpha,D)dD$. Both figures consider $\sigma_0/d_0^{a}=1$, $\alpha_p=1$ and $\alpha \in (0,2]$. }
\label{Fig6}
\end{figure}

\begin{figure}[h!]
\centering
\includegraphics[width=0.8\textwidth]{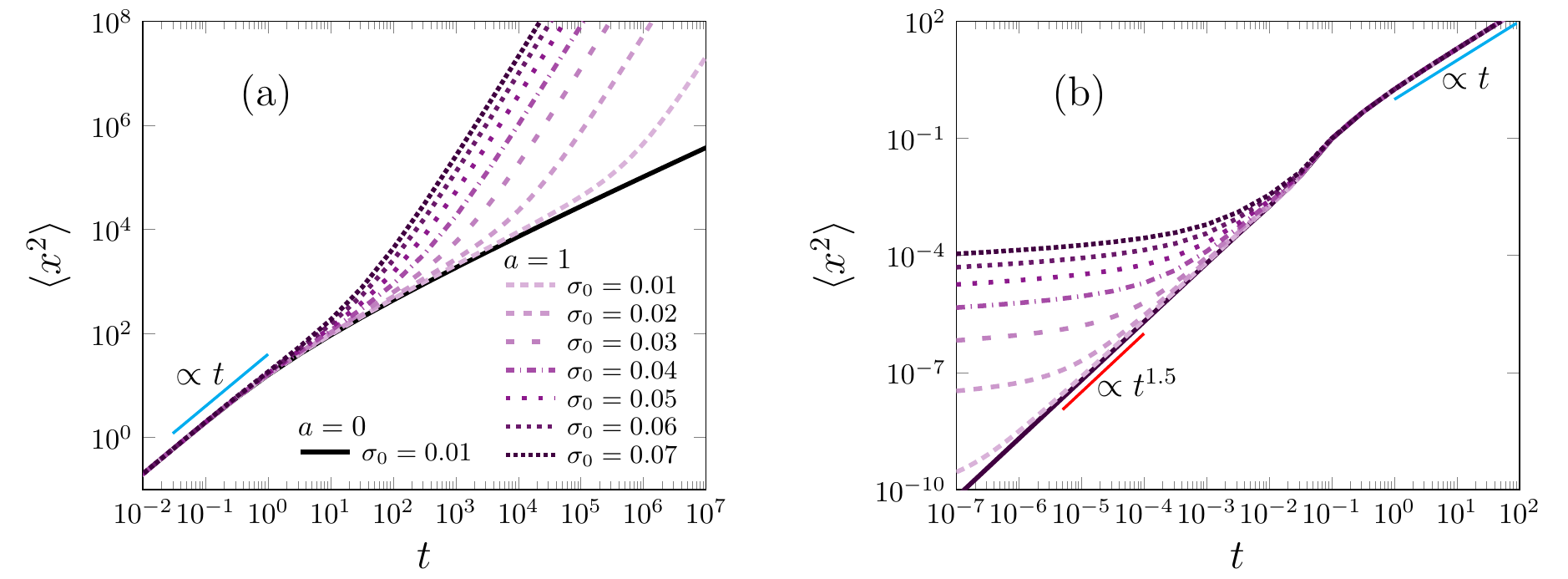}
\caption{Figures show MSD for different temporal scale but with same values of $\sigma_0$ in the PDF defined by Eq. (\ref{eq:GassianDalpha}). Fig. (a)  considers $\psi_{\alpha}(t)=  \left(1+t/\tau\right)^{\alpha}\log\left(1+t/\tau \right)$ into Eq. (\ref{eq:overallMSD2}),  in which $\tau=0.1$ and $\alpha=0.5$. Fig. (b)  considers $\psi_{\alpha}(t)= te^{-\tau/t} + t^{\alpha}e^{-t/\tau}$ into Eq. (\ref{eq:overallMSD2}), in which $\tau=0.1$ and $\alpha=1.5$. Both figures consider same values for $a=1$, $D_0=1$, $d_0=1$, $\alpha_p=1$ and  $\alpha \in (0,2]$. Particularly, the dashed curves represent the uncoupled situations between $\alpha$ and $D$, i.e. $a=0$ in Eq. (\ref{eq:coupled}).}
\label{Fig7}
\end{figure}

Fig. (\ref{Fig6}) complements our analysis by showing us the effects of coupling the variable $a$ over the effective diffusivity and distribution of anomalous exponents, in Figs. (\ref{Fig6}-a) and (\ref{Fig6}-b), respectively. These results show how the choices of diffusivity values can change the choices of the anomalous diffusion; the $(\alpha,D)$-coupling may be explored for other superstatistics that are connected with different experimental scenarios.

With this background of a conditioned PDF for $\alpha$ and $D$,  Fig. (\ref{Fig7}) shows different temporal scales for $\psi_{\alpha}$. In Fig. (\ref{Fig7}-a) and (\ref{Fig7}-b), the superstatistics on the anomalous exponent is relevant for long times and short times, respectively. Thereby, Fig. (\ref{Fig7}-a) and Fig. (\ref{Fig7}-b) reveal that the coupling parameter $a=1$ presents a strong break of power-law time relation in MSD behaviour for $\sigma_0 \sim 0.07$ ($\sigma \simeq 0.07 D$) in long and short temporal scales, respectively. Moreover, there is an evident crossover between short and long times in which, for $\sigma_0=0.01$, we may observe a power-law behaviour.

\newpage

\section{Conclusion}
\label{sec5}

Ensembles of anomalous tracers have been reported in different experimental settings and can be associated with different heterogeneous complex systems.  More recently, the two-variable superstatistics of fractional Brownian walkers was proposed to describe protein diffusion in bacteria \cite{BeckItto}. Further, current experimental investigations with single-particle tracking technique in tracers have reported statistics of anomalous exponents and diffusivities, for instance in endosomes trajectories inside living eukaryotic cells \cite{korabel2021local}, spreading dynamics of amoeboid cells \cite{cherstvy2018non}, diffusion of tracers in mucin hydrogels \cite{cherstvy2019non} and beads diffusion in Xenopus extracts \cite{speckner2021single}. In this framework, our investigation promotes advances in theoretical models that reveal anomalous non-Gaussian diffusion as a macroscopic behaviour which is a consequence of superstatistics of the individual anomalous tracers.

We have discussed the current scenarios that embrace the PDF of the anomalous diffusion exponents and diffusion coefficients for tracers diffusing in complex environments. Then, we proposed two different truncated superstatistics for the $\alpha$-anomalous exponent to apply in scaled Brownian motion. Numerous anomalous exponent distributions in literature are contained in the considered cases, which are truncated Gaussian and $\chi^2$-Gamma distributions. Firstly, the situations without diffusivity effects were considered, i.e. $f(D)\simeq \delta(D-D_0)$, so that it does not overshadow the effects of the anomalous exponent. Thereby, we have realised numerical simulations of heterogeneous scaled Brownian equation to present a match with the superstatistical model. This allowed us to explore the PDF in  space and MSD in time, which presents non-Gaussian distribution without the effects of diffusivities.

Furthermore, we also have considered other temporal scaling $\psi(t)$ in sBm, to obtain power-law time behaviours for MSD. Particularly, we discuss why a logarithmic time factor is necessary in temporal scaling to obtain an effective anomalous exponent for long time behaviours.  We have also described how the heterogeneous tracers are affected when the truncated Gaussian PDF for the anomalous exponent is conditioned to diffusivity values, i.e. $\pi(\alpha,D)\simeq f(D)\Tilde{g}(\alpha|D)$.  This investigation opens new interesting approaches that may be useful for describing particles diffusing in disordered environments. 
A large diversity of new situations can be approached for different $\alpha$ distributions as well as for other temporal scales $\psi_{\alpha}(t)$ in sBm.

\section*{Acknowledgements} 

We thank the Coordenação de Aperfeiçoamento de Pessoal de Nível Superior (CAPES), Grant No.  001.

\end{document}